\documentstyle[12pt, epsf]{article}

\advance \topmargin by -\headheight
\advance \topmargin by -\headsep

\evensidemargin \oddsidemargin
\def\ie{{\em i.e.}}

\def\CC{{\mathchoice
{\rm C\mkern-8mu\vrule height1.45ex depth-.05ex 
width.05em\mkern9mu\kern-.05em}
{\rm C\mkern-8mu\vrule height1.45ex depth-.05ex 
width.05em\mkern9mu\kern-.05em}
{\rm C\mkern-8mu\vrule height1ex depth-.07ex 
width.035em\mkern9mu\kern-.035em}
{\rm C\mkern-8mu\vrule height.65ex depth-.1ex 
width.025em\mkern8mu\kern-.025em}}}

\def\RR{{\rm I\kern-1.6pt {\rm R}}}

\def\ZZ{{\rm Z}\kern-3.8pt {\rm Z} \kern2pt}

\def\np{Nucl. Phys.}
\def\pl{Phys. Lett.}

\def\jmp{J. Math. Phys.}
\def\ijmp{Int. J. Mod. Phys.}
\def\mpl{Mod. Phys. Lett.}

\def\pnas{Proc. Natl. Acad. Sci. USA}

\def\qjm{Quart. J. Math. Oxford Ser. (2)}

\newcommand{\beq}{\begin{equation}}
\newcommand{\eeq}{\end{equation}}
\newcommand{\rc}{\nonumber\\}
\newcommand{\bear}{\begin{eqnarray}}
\newcommand{\eear}{\end{eqnarray}}

\newfont{\namefont}{cmr10}
\newfont{\addfont}{cmti7 scaled 1440}
\newfont{\boldmathfont}{cmbx10}
\newfont{\headfontb}{cmbx10 scaled 1728}
\begin{document}
\begin{titlepage}

\begin{center} \Large \bf 
osp$\bf{(1\vert 2)}$ Conformal Field Theory 

\end{center}

\vskip 0.3truein
\begin{center} 
I.P. Ennes
\footnote{e-mail:ennes@gaes.usc.es} 
A.V. Ramallo
\footnote{e-mail:alfonso@gaes.usc.es} 
and 
J. M. Sanchez de Santos
\footnote{e-mail:santos@gaes.usc.es}

\vspace{0.3in}

Departamento de F\'\i sica de
Part\'\i culas, \\ Universidad de Santiago\\
E-15706 Santiago de Compostela, Spain.

\end{center}
\vskip 1truein

\begin{center}
\bf ABSTRACT
\end{center} 

We review some results recently obtained for the
conformal field theories based on the affine Lie
superalgebra ${\rm osp}(1\vert 2)$. In particular, we
study the representation theory of the 
${\rm osp}(1\vert 2)$ current algebras and their
character formulas. By means of a free field
representation of the conformal blocks, we obtain the
structure constants and the fusion rules of the model. 
\vskip 1truein
{\it Lecture delivered at the CERN-Santiago de
Compostela-La Plata Meeting, ``Trends in Theoretical
Physics", La Plata, Argentina, April-May 1997.}

\vskip3.5truecm
\leftline{US-FT-26/97 \hfill August 1997}
\leftline{hep-th/9708094}
\smallskip
\end{titlepage}
\setcounter{footnote}{0}

\setcounter{equation}{0}
\section{Introduction and motivation}
\label{ospintro}

Among the two-dimensional theories endowed with
conformal invariance, those which, in addition, 
possess a current algebra symmetry are specially
important \cite{Review}. In this lecture, 
we shall report on some results we have recently
obtained for Conformal Fields Theories (CFT's) which
enjoy an ${\rm osp}(1\vert 2)$ affine superalgebra. As a
motivation for the study of this particular case, let us
mention that the 
${\rm osp}(1\vert 2)$ super Lie algebra has an
ubiquitous presence in many problems in which 
the $N=1$ superconformal symmetry is involved. Indeed,
the minimal $N=1$ superconformal models can be obtained
by means of Hamiltonian reduction of a system with 
${\rm osp}(1\vert 2)$ current algebra and this
symmetry appears in the light-cone approach to
two-dimensional supergravity 
\cite{bershadsky,poly}. It is also interesting to
mention in this respect that the topological 
${\rm osp}(1\vert 2)/{\rm osp}(1\vert 2)$ coset
theories can be used to describe the non-critical 
Ramond-Neveu-Schwarz superstrings \cite{yu,Ennes}. As
a final motivation let us point out that, as will be
shown below, a lot of non-trivial results can be found
for the ${\rm osp}(1\vert 2)$ CFT's. These results can be
simply stated and compared with those corresponding to 
CFT's based on the su$(2)$ affine Lie algebra.

The organization of this
 lecture is the following. In
section \ref{osprep} we recall the basic facts of the 
${\rm osp}(1\vert 2)$ representation theory. Its
similarity with ordinary angular momentum theory will
become evident and will constitute a guiding principle
for what follows. The ${\rm osp}(1\vert 2)$ current
algebra is introduced in section \ref{ospcurr} and the
corresponding character formulas are analyzed in
section \ref{ospchar}. In section \ref{ospfree} we study
a representation of the affine ${\rm osp}(1\vert 2)$
symmetry in terms of free fields. This representation
can be used to give integral expressions for the
conformal blocks, from which the structure constants
and  the fusion rules of the model can be extracted.
Finally, in section \ref{ospcon} some conclusions are
drawn and a series of final remarks are made.

\setcounter{equation}{0}
\section{ osp$\bf{(1\vert 2)}$ Representation Theory }
\label{osprep}

The ${\rm osp}(1\vert 2)$ superalgebra is a graded 
extension of the $sl(2)$ Lie algebra \cite{Pais}. It is
generated by three bosonic generators ($T_3$ and
$T_{\pm}$ ) and by two fermionic operators ($F_{\pm}$).
The bosonic generators close an $sl(2)$ algebra. The
full set of (anti)commutators that define the ${\rm
osp}(1\vert 2)$  superalgebra is:
\bear
&&[T_3\,,\,T_{\pm}]\,=\,\pm T_{\pm}
\,\,\,\,\,\,\,\,\,\,\,\,\,\,\,\,\,\,\,\,\,
\,\,\,\,\,\,\,\,\,\,\,\,\,\,\,\,\,\,\,\,\,
[T_{+}\,,\,T_{-}]\,=\,2T_3\rc
&&[T_3\,,\,F_{\pm}]\,=\,\pm {1\over 2} F_{\pm}
\,\,\,\,\,\,\,\,\,\,\,\,\,\,\,\,\,\,\,\,\,
\,\,\,\,\,\,\,\,\,\,\,\,\,\,\,\,
\{F_{\pm}\,,\,F_{\pm}\}\,=\,\pm 2 T_{\pm}
\label{uno}\\
&&\{F_{+}\,,\,F_{-}\}\,=\, 2 T_{3}
\,\,\,\,\,\,\,\,\,\,\,\,\,\,\,\,\,\,\,\,\,
\,\,\,\,\,\,\,\,\,\,\,\,\,\,\,\,\,\,\,\,\,
[T_{\pm}\,,\,F_{\pm}]\,=\,0\rc
&&[T_{\pm}\,,\,F_{\mp}]\,=\,-F_{\pm}\,\,.
\nonumber
\eear
It can be easily checked from (\ref{uno}) that the
operator:
\beq
C_2\,=\,T_3^2\,+\,{1\over 2}\,[\,T_-T_+\,+\,T_+T_-\,]\,+\,
{1\over 4}\,[\,F_-F_+\,-\,F_+F_-\,]\,,
\label{dos}
\eeq
commutes with all the generators of the 
${\rm osp}(1\vert 2)$ algebra. $C_2$ is the so-called
quadratic Casimir operator. Using the algebra defining
relations (eq. (\ref{uno})) one can reexpress $C_2$ as:
\beq
C_2\,=\,T_3^2\,+\,{1\over 2}\,T_3\,+\,T_-T_+\,+\,
{1\over 2}\,F_-F_+\,.
\label{tres}
\eeq
Following the standard methods of angular momentum theory,
one can find matrix representations of the algebra
(\ref{uno}). The finite dimensional
irreducible representations 
${\cal R}_{j}$ of the ${\rm osp}(1\vert 2)$ theory are
labeled by an integer or half-integer number $j$, which
we shall refer to as the isospin of the representation.
The highest weight vector of the representation 
${\cal R}_{j}$ will be denoted by $|j,j>$. It satisfies
the conditions:
\beq
T_+\,|j,j>\,=\,F_+\,|j,j>\,=\,0\,.
\label{cuatro}
\eeq
From the vector $|j,j>$,  one can easily obtain other
vectors of ${\cal R}_{j}$ by acting with the lowering
operators $T_-$ and $F_-$. We shall denote by $|j,m>$ to a
general basis state for the representation ${\cal R}_{j}$,
$m$ being the $T_3$ eigenvalue. The quadratic Casimir
operator $C_2$ acts on the states $|j,m>$ as a multiple of
the identity operator. The precise action of $C_2$ on the
states of  ${\cal R}_{j}$ can be determined by computing 
$C_2\,|j,j>$ from the highest weight conditions 
(\ref{cuatro}). The result is:
\beq
C_2\,|j,m>\,=\,j\,(\,j+{1\over 2}\,)\,|j,m>\,.
\label{cinco}
\eeq
It is not difficult to obtain the matrix elements of the
generators of ${\rm osp}(1\vert 2)$ in the representation 
${\cal R}_{j}$. For the bosonic generators one has:
\bear
&&T_{3}\,|j,m>\,=\,m|j,m>\rc\rc
&&T_{\pm}\,|j,m>\,=\,\sqrt{\,[j\mp m]\,[j\pm m+1]}\,
|j,m\pm 1>\,\,,
\label{seis}\\
\nonumber
\eear
where $[x]$ represents the integer part of the number $x$
($2x\in \ZZ$). The action of the operators $F_{\pm}$ on
the states $|j,m>$ is the following:
\beq
F_{\pm}\,|j,m>\,=\,\cases{
-\sqrt{j\mp m}\,|j,m\pm {1\over 2}>&if $j-m\in \ZZ$\cr\cr
\mp\sqrt{j\pm m+{1\over 2}}\,|j,m\pm {1\over 2}>
&if $j-m\in\ZZ+{1\over 2}$.}
\label{siete}
\eeq
Notice that the operators $T_{\pm}$ ($F_{\pm}$) change the 
$T_3$ eigenvalue in $\pm 1$ ($\pm 1/2$). In addition, the
fermionic operators change the statistics of the states.
It is clear from (\ref{seis}) and (\ref{siete}) that, 
when  $2j\in\ZZ$,  the representation ${\cal R}_j$ is 
$4j+1$-dimensional and spanned 
by the states $|j,m>$ with 
$m=-j, -j+{1\over 2}, \cdots, j-{1\over 2}, j$. In order
to characterize completely the representation one must
give the statistics of its highest weight state. The
Grassmann parity of $|j,j>$ will be denoted by $p(j)$ 
($p(j)=0,1$). We will say that the representation 
${\cal R}_j$ is even(odd) when $|j,j>$ is
bosonic(fermionic), \ie\ when $p(j)=0$($p(j)=1$). It is
clear that the Grassmann parity of the state $|j,m>$ is 
$p(j)+2(j-m)\,\,$ ${\rm mod}\,(2)$.

For Lie superalgebras, one can define a generalized 
adjoint operation, denoted by $\ddagger$,  such that,  
for any operator $A$ and any two states $\alpha$ and
$\beta$, one has:
\beq
<A^\ddagger \alpha | \beta>\,=\,(-1)^{p(A)p(\alpha)}\,\,
<\alpha|A\beta>\,.
\label{ocho}
\eeq

We shall call $A^\ddagger$ the superadjoint of $A$. In eq. 
(\ref{ocho}), $p(A)$ and $p(\alpha)$ denote respectively
the Grassmann parities of the operator $A$ and the state
$\alpha$. One can verify that the superadjoint of the
product of two operators is given by the formula:
\beq
(AB)^{\ddagger}\,=\,(-1)^{p(A)p(B)}\,\,
B^{\ddagger}\,A^{\ddagger}\,\,.
\label{nueve}
\eeq
It is easy to prove that the compatibility of the
property (\ref{nueve}) and the relations (\ref{uno})
requires that $T_{\pm}^{\,\ddagger}\,=\,T_{\mp}$ and 
$T_3^{\ddagger}\,=\,T_3$. In the case of the fermionic
generators we have, however, some freedom. Actually, if
$\eta$ is a number that can take the values $\pm 1$, the
rule which makes  the superadjoint $F_{\pm}^{\ddagger}$ 
consistent with the (anti)commutators (\ref{uno})
is:
\beq
F_{+}^{\ddagger}\,=\,\eta\,F_{-}
\,\,\,\,\,\,\,\,\,\,\,\,\,\,\,\,\,\,\,\,\,
F_{-}^{\ddagger}\,=\,-\eta\,F_{+}\,\,.
\label{diez}
\eeq
It is important to point out that,  in any case,  
$(\,(\,F_{\pm}\,)^{\ddagger}\,)^{\ddagger}\,=\,-F_{\pm}$.
The value of $\eta$ is related to the norm of the states
and the parity $p(j)$ of the representation. To illustrate
this point let us suppose that 
$<j,m|j,m>\,=\,\epsilon (\epsilon')$ if $j=m$ is
integer(half-integer), where $\epsilon$ and $\epsilon'$ can
take the value $+1$ or $-1$. Putting in eq. (\ref{ocho}) 
$\alpha\,=\,|j,j>$, $\beta\,=\,|j,j-{1\over 2}>$ and
$A\,=\,F_{+}$, one gets:
\beq
\eta\,=\,(-1)^{p(j)}\,\epsilon\epsilon\,'\,\,.
\label{once}
\eeq
Once we conventionally fix $\eta$ to a given value, the
signs $\epsilon$ and $\epsilon'$ of the norms of the
states are related to the highest weight parity $p(j)$ by
means of eq. (\ref{once}). For simplicity,  we shall
choose
$\eta=1$, which implies that 
$\epsilon\epsilon\,'\,=\,(-1)^{p(j)}$. For even
representations,  $p(j)=0$ and the norms $\epsilon$ and 
$\epsilon\,'$ can be taken to be $+1$, whereas for odd
representations,  $\epsilon$ and $\epsilon\,'$ must have
opposite sign. We shall choose 
$\epsilon\,=\,-\epsilon\,'\,=\,+1$ for odd representations
and, therefore, the norms of the states will be given by
the expression:
\beq
<j,m|j,m>\,=\,(-)^{2p(j)(j-m)}\,\,.
\label{doce}
\eeq

When two representations of isospins $j_1$ and $j_2$ are
coupled, one can decompose the corresponding tensor product
in the following way:
\beq
{\cal R}_{j_1}\,\otimes {\cal R}_{j_2}\,=\,
\bigoplus_{{j_3=|j_1-j_2|\atop}\atop 2(j_3-j_1-j_2)\,\in\,\ZZ}
^{j_1+j_2}
\,{\cal R}_{j_3}\,\,,
\label{trece}
\eeq
which means that one gets representations of isospins 
$|j_1-j_2|\,,\,|j_1-j_2|+{1\over 2}\,,\cdots, 
j_1+j_2-{1\over 2}\,,\,j_1+j_2$. The parity of the
representation ${\cal R}_{j_3}$ in the right-hand side of
eq. (\ref{trece}) is given by:
\beq
p(j_3)\,=\,p(j_1)\,+\,p(j_2)\,+\,2(j_1+j_2-j_3)
\,\,\,\,\,\,\,\,\,\,\,\,\,\,\,\,\,\,\,\,
{\rm mod}\,(2)\,\,.
\label{catorce}
\eeq
Notice that eq. (\ref{catorce}) implies that odd
representations appear when even representations are
coupled. In fact, if we denote by $[j]$ and 
$\widetilde {[j]}$ to the even and odd representations of
isospin $j$, eqs. (\ref{trece}) and (\ref{catorce}) imply,
in particular, that:
\bear
[{\scriptstyle{1\over 2}}]\,
\otimes [{\scriptstyle{1\over
2}}]\,=&&\,[0]+
\widetilde{[ {\scriptstyle{1\over 2}}]}+[1]\,\rc
\, [1]\,\otimes \,[1]\,=&&\,[0]+
\widetilde{[{\scriptstyle{1\over 2}}]}+[1]+
\widetilde{[{\scriptstyle {3\over 2}}]}+[2]\,\,.
\label{quince}\\
\nonumber
\eear
The presence of odd representations in the right-hand side
of eq. (\ref{quince}) means that one cannot avoid having
negative norm states and, therefore, a theory enjoying 
this symmetry cannot be unitary.

\setcounter{equation}{0}
\section{ osp$\bf{(1\vert 2)}$  Current Algebra }
\label{ospcurr}

In order to construct a Conformal Field Theory endowed
with the ${\rm osp}(1\vert 2)$ symmetry, one must first
extend the finite algebra of section \ref{osprep} to the
affine, infinite dimensional, ${\rm osp}(1\vert 2)$ Lie
superalgebra. As it is well-known, this can be achieved by
replacing the generators of section \ref{osprep} by
currents depending on a holomorphic variable $z$:
\bear
&&T_{\pm}\,\,\Longrightarrow\,\,
J^{\pm}(z)\,=\,\sum_{n=-\infty}^{+\infty}
J^{\pm}_n\,z^{-n-1}\rc
&&T_{3}\,\,\Longrightarrow\,\,
J^{0}(z)\,=\,\sum_{n=-\infty}^{+\infty}
J^{0}_n\,z^{-n-1}\label{dseis}\\
&&F_{\pm}\,\,\Longrightarrow\,\,
j^{\pm}(z)\,=\,\sum_{n=-\infty}^{+\infty}\,\,
j^{\pm}_n\,z^{-n-1}\,\,.\rc
\nonumber
\eear
In eq. (\ref{dseis}), we have displayed the mode
expansions of the different currents. Notice that the
modes $n$ of the fermionic currents run over the
integers, which implies that we are considering the
Ramond sector of the 
${\rm osp}(1\vert 2)$ affine superalgebra. The
non-vanishing (anti)commutators of the currents 
$J_n^a$ and $j_n^{\alpha}$ are:
\bear
&&[\,J_n^0\,,\,J_m^{\pm}\,]\,=\,\pm J_{n+m}^{\pm}
\,\,\,\,\,\,\,\,\,\,\,\,\,\,\,\,\,\,\,\,\,\,\,\,\,
\,\,\,\,\,\,\,\,\,\,\,\,\,\,\,\,\,\,\,\,\,\,\,\,\,
[\,J_n^0\,,\,J_m^{0}\,]\,=\,
{k\over 2}\,n\,\delta_{n+m}\rc
&&[\,J_n^+\,,\,J_m^{-}\,]\,=\,kn\delta_{n+m}\,+\,
2J^0_{n+m}\rc
&&[\,J_n^0\,,\,j_m^{\pm}\,]\,=\,\pm\,{1\over 2}\,
j_{m+n}^{\pm}
\,\,\,\,\,\,\,\,\,\,\,\,\,\,\,\,\,\,\,\,\,\,\,\,\,
\,\,\,\,\,\,\,\,\,\,\,\,\,\,\,\,\,\,\,\,
[\,J_n^{\pm}\,,\,j_m^{\pm}\,]\,=\,0\label{dsiete}\\
&&[\,J_n^{\pm}\,,\,j_m^{\mp}\,]\,=\,-j_{n+m}^{\pm}
\,\,\,\,\,\,\,\,\,\,\,\,\,\,\,\,\,\,\,\,\,\,\,\,\,
\,\,\,\,\,\,\,\,\,\,\,\,\,\,\,\,\,\,\,\,\,\,\,\,\,
\{\,j_n^{\pm}\,,\,j_m^{\pm}\,\}\,=
\,\pm 2 J_{n+m}^{\pm}\rc
&&\{\,j_n^+\,,\,j_m^{-}\,\}\,=\,2kn\delta_{n+m}\,+\,
2J^0_{n+m}\,\,.\rc
\nonumber
\eear
In what follows, the algebra defined in (\ref{dsiete}) 
will be simply denoted by ${\cal A}$. By inspecting eq. 
(\ref{dsiete}),  one can verify that the zero modes 
$J_0^a$ and $j_0^{\alpha}$ of the currents close the
algebra (\ref{uno}). In eq. (\ref{dsiete}),  $k$ is a
central element (the level of the ${\rm osp}(1\vert 2)$ 
current algebra) which commutes with all the other
generators. By means of the Sugawara prescription, one
can  construct an energy-momentum tensor $T(z)$ for the 
${\rm osp}(1\vert 2)$  currents. The expression of $T(z)$
is the following:
\bear
T(z)\,=\,{1\over 2k+3}\,:\,[\,&&2\,(J^0(z))^2\,+\,
J^+(z)\,J^-(z)\,+\,
J^-(z)\,J^+(z)\,-\nonumber\\
&&-\,{1\over 2}\,j^+(z)\,j^-(z)\,+\,
{1\over 2}\,j^-(z)\,j^+(z)\,]:\,\,,
\label{docho}\\\nonumber
\eear
where the double dot $:$ denotes normal ordering. The
modes $L_n$ of the energy-momentum tensor are defined as:
\beq
T(z)\,=\,\sum_{n=-\infty}^{+\infty}\,\,
L_n\,z^{-n-2}\,\,.
\label{dnueve}
\eeq
A calculation performed with the standard techniques of
CFT allows to prove that the commutators of the $L_n$'s
with the currents are:
\beq
[\,L_n\,,\,J_m^a\,]\,=\,-m\,J_{n+m}^a
\,\,\,\,\,\,\,\,\,\,\,\,\,\,\,\,\,\,\,\,
[\,L_n\,,\,j_m^{\alpha}\,]\,=\,-m\,j_{n+m}^{\alpha}\,\,.
\label{veinte}
\eeq
Similarly, one can verify that the modes of the
energy-momentum tensor satisfy the Virasoro algebra:
\beq
[\,L_n,L_m\,]\,=\,(n-m)\,L_{n+m}\,+\,{c\over 12}\,\,
(m^3\,-\,m)\,\delta_{n+m,0}\,\,,
\label{vuno}
\eeq
where the central charge $c$ is related to the level $k$
by means of the expression:

\beq
c\,=\,{2k\over 2k+3}\,\,.
\label{vdos}
\eeq

In the algebra (\ref{dsiete}), we can introduce the
so-called principal gradation, which is defined as:
\beq
d(\,J_n^a\,)\,=\,2n\,+\,a
\,\,\,\,\,\,\,\,\,\,\,\,\,\,\,
d(\,j_n^{\alpha}\,)\,=\,2n\,+\,{\alpha\over 2}
\,\,\,\,\,\,\,\,\,\,\,\,\,\,\,
d(\,k\,)\,=\,0\,\,.
\label{vtres}
\eeq
With respect to $d$, the algebra ${\cal A}$ splits as:
\beq
{\cal A}\,=\,{\cal A}_-\,\oplus\,{\cal A}_0\,
\oplus\,{\cal A}_+\,\,,
\label{vcuatro}
\eeq
where ${\cal A}_-$,  ${\cal A}_0$ and  ${\cal A}_+$ are
the subspaces of  ${\cal A}$ spanned by the elements that
have, respectively, $d<0$, $d=0$ and $d>0$. These
elements are easy to identify from eq. (\ref{vtres}) and so,
for example,  ${\cal A}_0$ is generated by $J_0^0$ and
$k$, whereas  ${\cal A}_+$ is the subspace spanned by 
$J_n^-\,\,(n\ge 1)$, $J_n^0\,\,(n\ge 1)$, 
$J_n^+\,\,(n\ge 0)$, $j_n^-\,\,(n\ge 1)$ and 
$j_n^+\,\,(n\ge 0)$.

The Verma modules associated to ${\cal A}$ are
constructed by acting with elements of the universal
enveloping algebra of ${\cal A}_-$ (denoted by 
$U(\,{\cal A}_-\,)$) on a highest weight vector $|\,j,k>$.
The latter is annihilated by the elements of  
${\cal A}_+$, \ie:
\beq
J_n^a\,|\,j,k>\,=\,j_n^{\alpha}\,|\,j,k>\,=\,0\,\,,
\,\,\,\,\,\,\,\,\,\,\,\,\,\,\,\,\,\,
\forall\,\,(\,J_n^a\,,\,j_n^{\alpha}\,)\,\in\,
{\cal A}_{+}\,\,.
\label{vcinco}
\eeq
On the contrary, $J_0^0$ and $L_0$ act diagonally on 
$|\,j,k>$:
\beq
J_0^0\,|\,j,k>\,=\,j\,|\,j,k>
\,\,\,\,\,\,\,\,\,\,\,\,\,\,\,\,\,\,
L_0\,|\,j,k>\,=\,h_j\,|\,j,k>\,\,.
\label{vseis}
\eeq
From the Sugawara expression for $L_0$ (see eqs.
(\ref{docho}) and (\ref{dnueve}) ), one can easily get the
$L_0$ eigenvalue corresponding to $|\,j,k>$, namely:
\beq
h_j\,=\,{j\,(\,2j\,+\,1\,)\over 2k\,+\,3}\,\,.
\label{vsiete}
\eeq
As in the case of the finite algebra, 
in order to characterize completely   the highest
weight vector $|\,j,k>$,  we must specify its Grassmann
parity, which we shall also denote by $p(j)$. The Verma
module whose highest weight vector is $|\,j,k>$ will be
denoted by $V^{(j,k)}$. Any element in $V^{(j,k)}$ is of
the form 
$u_-|\,j,k>$, where $u_-\in {\cal A}_-$. Notice that,
according to the Poincar\'e-Birkhoff-Witt theorem,
$U(\,{\cal A}_-\,)$ is generated by monomials and thus we
can consider a basis of $V^{(j,k)}$ constituted by vectors
of the form: 
\beq
{|\,\{m_i^a\}\,;j\,>}\,=\,
\prod_{i=0}^{+\infty}\,\Bigl(\,j_{-i}^-\,\Bigr)^{2m_i^-}\,\,
\prod_{i=1}^{+\infty}\,\Bigl(\,J_{-i}^0\,\Bigr)^{m_i^0}\,\,
\prod_{i=1}^{+\infty}\,\Bigl(\,j_{-i}^+\,\Bigr)^{2m_i^+}\,\,
|\,j,k>\,\,.
\label{vocho}
\eeq
In eq. (\ref{vocho}), the numbers 
$m_i^{\pm}$ are integers or half-integers whereas the 
$m_i^0$'s are always integers ($m_i^a\ge 0$).

For some values of the isospin $j$ the Verma module 
$V^{(j,k)}$ is reducible, \ie\ it contains singular
vectors. These are vectors of $V^{(j,k)}$  which are
descendants and are annihilated by ${\cal A}_+$. For a
given value of the level $k$, the singular vectors appear
in  those modules with highest weight vectors whose
isospins belong to a discrete set labelled by two integers
$r$ and $s$. These isospins are of the form \cite{KW}:
\beq
4j_{r,s}\,+\,1\,=\,r\,-\,s\,(\,2k+3\,)\,\,,
\label{vnueve}
\eeq
where $r+s$ is odd and, either $r>0$ and $s\ge 0$ or
$r<0$ and $s<0$. The $J_0^0$ and $L_0$ eigenvalues of
these vectors are respectively $j_{r,s}\,-{r\over 2}$
and 
$h_{j_{r,s}}\,+\,{rs\over 2}$.

\setcounter{equation}{0}
\section{ osp$\bf{(1\vert 2)}$ character formulae }
\label{ospchar}

Let us now study the characters of the 
${\rm osp}(1\vert 2)$ CFT. For an irreducible Verma module 
$V^{(j,k)}$, whose highest weight vector has isospin $j$,
the characters $\lambda_{j}(a,\tau)$ are defined as:
\beq
\lambda_{j}(a,\tau)\,=\,{\rm Tr}_j\,
[\,q^{L_0-{c\over 24}}\,w^{J_0^0}\,]\,,
\label{treinta}
\eeq
where the trace is taken over the module $V^{(j,k)}$ and
$q$ and $w$ are two variables related to the modular
parameter $\tau$ and to the Cartan coordinate $a$ by means
of the expressions:
\beq
q\,=\,e^{2\pi i \tau}
\,\,\,\,\,\,\,\,\,\,\,\,\,\,\,\,\,\,\,\,
w\,=\,e^{2\pi i a}\,\,.
\label{tuno}
\eeq
The trace in eq.  (\ref{treinta})  can be evaluated by
studying the action of the operator 
$q^{L_0-{c\over 24}}\,w^{J_0^0}$ on the states 
$|\,\{m_i^a\}\,;j\,>$ defined in eq. (\ref{vocho}). Since
$L_0$ and $J_0^0$ act diagonally on these states, the
trace (\ref{treinta}) can be easily calculated. After
some simple manipulations \cite{KW,yudos}, one obtains
the following expression for $\lambda_{j}(a,\tau)$:
\beq
\lambda_{j}(a, \tau)\,=\,
{q^{ {2(j\,+{1\over 4})^2\over 
2k\,+\,3}}\,w^{j+\,{1\over 4}}
\over\Pi(a,\tau)}\,,
\label{tdos}
\eeq
where the function $\Pi(a, \tau)$, appearing in the
denominator,  is the following infinite product:
\beq
\Pi(a, \tau)\,\equiv\,q^{{1\over 24}}\,w^{{1\over 4}}\,
\prod_{n=1}^{+\infty}\,\,
(1-q^n)\,(1-w^{{1\over 2}}q^n)\,(1-w^{-{1\over 2}}q^{n-1})
\,(1-wq^{2n-1})\,(1-w^{-1}q^{2n-1})\,.
\label{ttres}
\eeq
By means of the Watson quintuple product identity:
\bear
\prod_{n=1}^{+\infty}\,\,&&
(1-q^n)\,(1-wq^n)\,(1-w^{-1}q^{n-1})
\,(1-w^2q^{2n-1})\,(1-w^{-2}q^{2n-1})\,=\,\rc
&&=\,\sum_{m=-\infty}^{+\infty}\,\,
(w^{3m}-w^{-3m-1})\,q^{{3m^2+m\over 2}}\,\,,
\label{tcuatro}\\
\nonumber
\eear
one can write $\Pi(a, \tau)$ in the form:
\beq
\Pi(a,\tau)\,=\,
\Theta_{1,3}\,( {a\over 2},{\tau\over 2})\,-\,
\Theta_{-1,3}\,({a\over 2},{\tau\over 2})\,\,,
\label{tcinco}
\eeq
where $\Theta_{r,s}$ are the classical theta functions,
defined as:
\beq
\Theta_{r,s}\,(a,\tau)\,=\,
\sum_{m\in \ZZ}\,q^{s(m+{r\over 2s})^2}\,
w^{s(m+{r\over 2s})}\,.
\label{tseis}
\eeq

For some particular values of the level $k$ there exists a
class of representations which 
are completely degenerate \cite{KW,yudos}.
These representations occur for values of $k$ which are
rational numbers of the form:
\beq
2k+3\,=\,{p\over p'}\,\,,
\label{tsiete}
\eeq
where $p$ and $p'$ are coprime positive integers such that
$p+p'$ is even and $p$ and $(p+p')/2$ are relatively
prime. The so-called admissible representations
correspond to isospins of the form:
\beq
4j_{r,s}\,+\,1\,=\,r\,-\,s\,{p\over p\,'}\,\,,
\label{tocho}
\eeq
with $r$ and $s$ taking values in the grid 
$1\le r\le p-1$, $0\le s\le p'-1$ and $r+s\in 2\ZZ+1$.
When the isospin is of the form (\ref{tocho}), the
corresponding Verma module will have a null vector, since 
eq. (\ref{tocho}) corresponds to eq. (\ref{vnueve}) with 
$r>0$ and $s\ge 0$. Moreover, when eqs. (\ref{tsiete}) and 
(\ref{tocho}) are satisfied, one has that 
$j_{r,s}\,=\,j_{r-p,s-p'}$ and, therefore, when $r$ and
$s$ belong to the grid defined above, the
isospin (\ref{tocho})  has also  the form (\ref{vnueve})
for the integers $r-p<0$ and $s-p'<0$. Therefore, when the
isospin $j_{r,s}$ belongs to the admissible set
(\ref{tocho}), the module $V^{j_{r,s}, k}$ possesses a
second singular vector. These two null vectors generate
the maximum proper submodule of $V^{j_{r,s}, k}$, which
can be generated by means of the embedding diagram:
$$
\def\sp{\nearrow\!\!\!\!\!\!\searrow}
\matrix{&&B(0)&\longrightarrow&B(1)&\longrightarrow&B(1)
&\longrightarrow&A(2)&\longrightarrow&\cdots\cr
A(0)&\nearrow\atop\searrow&&\sp&&\sp&&\sp&&\sp&\cdots\cr
&&B(-1)&\longrightarrow&A(-1)&\longrightarrow
&B(-2)&\longrightarrow&A(-2)
&\longrightarrow&\cdots\cr}
$$
where $A(l)$ and $B(l)$ are given by:
\bear
A(l)\,&&\equiv\,j_{r-2lp\,,\,s}\,=\,
{r-1\over 4}\,-\,{s\over 4}\,{p\over p\,'}\,-\,
l\,{p\over 2}\rc
B(l)\,&&\equiv\,j_{-r-2lp\,,\,s}\,=\,
{r-1\over 4}\,-\,{s\over 4}\,{p\over p\,'}\,-\,
l\,{p\over 2}\,-\,{r\over 2}\,\,.\label{tnueve}\\
\nonumber
\eear
Each node in the above diagram represents a Verma module
with $A(l)$ or $B(l)$ as the isospin of its highest
weight state. An arrow connecting two spaces 
$E\rightarrow F$ means that the module $F$ is contained
in the module $E$. The character of the irreducible
module with isospin $j=j_{r,s}$ is constructed as an
alternating sum of the form:
\beq
\chi_{j_{r,s}}(a,\tau)\,=\,
\sum_{l=-\infty}^{l=+\infty}\,\,
\lambda_{A(l)}(a,\tau)\,-\,
\sum_{l=-\infty}^{l=+\infty}\,\,
\lambda_{B(l)}(a,\tau)\,\,.
\label{cuarenta}
\eeq
Using eqs. (\ref{tdos})  and (\ref{tnueve})  in the
right-hand side of eq. (\ref{cuarenta}), 
it is straightforward to prove
that $\chi_{j_{r,s}}(a,\tau)$ can be written as a
quotient of differences of theta functions. Actually,
defining the constants $b_{\pm}$ and $e$ as:
\beq
b_{\pm}\,=\,\pm p\,' r\,-\,p\,s
\,\,\,\,\,\,\,\,\,\,\,\,\,\,\,\,\,\,\,\,\,\,\,\,\,
e\,=\,p\,p\,'\,\,,
\label{cuno}
\eeq
the characters $\chi_{j_{r,s}}(a,\tau)$ can be put in the
form:
\beq
\chi_{j_{r,s}}(a,\tau)\,=\,
{\Theta_{b_{+}, e}({a\over 2p\,'},{\tau\over 2}\,)\,-\,
\Theta_{b_{-}, e}( {a\over 2p\,'},{\tau\over 2}\,)
\over \Pi(a, \tau)}\,\,.
\label{cdos}
\eeq

It is interesting to study the behaviour of the
characters (\ref{cdos}) when $a\rightarrow 0$
\cite{null}. First of all, it is easy to prove that the
denominator 
$\Pi(a, \tau)$ vanishes linearly when 
$a\rightarrow 0$. Actually, one can check that:
\beq
\Pi(a,\tau)\,=\,i\pi a q^{{1\over 24}}\,
\sum_{m\in \ZZ}\,(6m+1)\,q^{{3m^2+m\over 2}}\,+\,
o(a^2)\,\,.
\label{ctres}
\eeq
In general, the numerator of the right-hand side
of eq. (\ref{cdos}) does not vanish when $a=0$.
Therefore $\chi_{j_{r,s}}(a,\tau)$ will, in
general, develop a single pole in $a$ in the 
$a\rightarrow 0$ limit. By studying the residue of
the  ${\rm osp}(1\vert 2)$ characters in this
singularity we are going to discover a remarkable
connection with the minimal supersymmetric models. Let
us, first of all, rewrite the infinite sum appearing in
the right-hand side of eq. (\ref{ctres}) as an infinite
product. An identity due to Gordon \cite{Gordon} states
that:
\beq
q^{{1\over 24}}\,
\sum_{m\in \ZZ}\,(6m+1)\,q^{{3m^2+m\over 2}}
\,=\,2\,
{\Big[\,\eta(\tau)\,\Big]^4\over
\theta_2(0,\tau)}\,\,,
\label{ccuatro}
\eeq
where $\eta (\tau)$ is the Dedekind $\eta$-function,
which can be represented as:
\beq
\eta (\tau)\,=\,q^{{1\over 24}}\,
\prod_{n=1}^{\infty}\,(1\,-\,q^n)\,\,.
\label{ccinco}
\eeq
and $\theta_2(0,\tau)$ is a Jacobi theta function,
whose infinite product representation can be obtained
from (\ref{ccinco}) and from the following relation
with  $\eta (\tau)$:
\beq
{\theta_2(0,\tau)\over \eta (\tau)}\,=\,
2\,q^{{1\over 12}}\,\prod_{n=1}^{\infty}\,
(\,1\,+\,q^{n}\,)^2\,\,.
\label{cseis}
\eeq
It is easy to verify that  the numerator of the 
${\rm osp}(1\vert 2)$ characters does not vanish when 
$s\not= 0$ (see eq. (\ref{cdos})). Therefore, it makes
sense to consider the residue of
$\chi_{j_{r,s}}(a,\tau)$ at the point
$a=0$. Let us define for $s\not= 0$ the following
quantity:
\beq
\hat\chi_{r,s}(\tau)\,\equiv\,
\Big[\,{2\eta(\tau)\over\theta_2(0,\tau)}\,
\Big]^{1\over 2}\,\,
\Big[\,\eta(\tau)\,\Big]^2\,\,
{\rm lim}_{a\rightarrow 0}\,\,
\Big\{\,i\pi z\,
\chi_{j_{r,s}}(a,\tau)
\,\Big\}\,\,.
\label{csiete}
\eeq
Using the Gordon identity  (\ref{ccuatro}),  one can
demonstrate that  $\hat\chi_{r,s}(\tau)$ is given by:
\beq
\hat\chi_{r,s}(\tau)\,=\,
\Big[\,{\theta_2(0,\tau)
\over 2\eta(\tau)}\,\Big]^{1\over 2}\,\,\,\,
{\Theta_{b_{+}, e}(\,0,{\tau\over 2}\,)\,-\,
\Theta_{b_{-}, e}(\,0,{\tau\over 2}\,)\over
\eta(\tau)}\,\,.
\label{cocho}
\eeq
It is interesting to point out that, for 
$1\le r\le p-1\,\,\,$, $1\le s\le p\,'-1\,\,$  and
$r+s\in 2\ZZ+1$, the functions of $\tau$ appearing in
the right-hand side of eq. (\ref{cocho}) are precisely
the characters of the minimal supersymmetric models, with
central charge 
$c\,=\,{3\over 2}\,(1\,-\,{2(p-p\,')^2\over pp\,'})$, in the
Ramond sector. This is precisely the result we were
looking for.

\setcounter{equation}{0}
\section{ Free field representation}
\label{ospfree}

The ${\rm osp}(1\vert 2)$ current algebra can be 
realized \cite{bershadsky,osp} in terms of free
fields. The field content of this representation consists
 of an scalar field $\phi$, a
pair of two conjugate bosonic field $(w, \chi)$ and two
fermionic fields $(\psi, \bar\psi)$ whose non-vanishing
operator expansions (OPE's) are:
\beq
w(z_1)\,\chi(z_2)\,=\,\psi(z_1)\,\bar\psi(z_2)\,=\,{1\over z_1-z_2}
\,\,\,\,\,\,\,\,\,\,\,\,\,\,\,\,\,\,
\phi(z_1)\,\phi(z_2)\,=\,-{\rm log}\,(z_1-z_2)\,.
\label{cnueve}
\eeq
In terms of these fields the expression of the currents
is:
\bear
J^+\,=&&\,w\rc
J^-\,=&&-\,w\chi^2\,+\,i\sqrt{2k+3}\,\,\chi\partial\phi\,-
\,\chi\psi\bar\psi\,+k\partial\chi\,+
\,(k+1)\psi\partial\psi\rc
J^0\,=&&-w\chi\,+{i\over
2}\,\sqrt{2k+3}\,\,\partial\phi\,-\, {1\over
2}\,\psi\bar\psi\label{cincuenta}\\ 
j^+\,=&&\bar\psi\,+\,w\psi\rc
j^-\,=&&-\chi(\bar\psi\,+\,w\psi)\,+i\sqrt{2k+3}\,\,
\psi\partial\phi\,+\,(2k+1)\partial\psi\,.\rc
\nonumber
\eear
Substituting eq. (\ref{cincuenta}) in the Sugawara
expression of $T$ (eq. (\ref{docho})), one gets:
\beq
T\,=\,w\partial\chi\,-\,\bar\psi\partial
\psi\,-\,{1\over 2}\,(\partial\phi)^2\,+\,
{i\over 2}\,\alpha_0\,\partial^2\phi\,,
\label{ciuno}
\eeq
where the background charge of the $\phi$ field is given
by:
\beq
\alpha_0\,=\,-{1\over \sqrt{2k+3}}\,.
\label{cidos}
\eeq

Let us now construct the primary fields of the model
\cite{osp}. The primary field associated to the state
$|j,m>$ of the representation ${\cal R}_j$ of the finite
algebra  (\ref{uno}) will be denoted by $\Phi^j_m$. In
what follows, we shall restrict ourselves to the case in
which the level $k$ is a positive integer. Notice that  
this corresponds to taking $p'=1$ in eq.
(\ref{tsiete}). Therefore, the isospins corresponding
to the admissible representations are given by eq. 
(\ref{tocho}) with $s=0$. As $r$ in eq. (\ref{tocho})
must be odd, the highest value it can take is $2k+1$
and, thus, we conclude that the admissible
representations have integer or half-integer isospins
$j$ that satisfy $j\le k/2$. It will be understood from
now on that this constraint is satisfied by all primary
fields $\Phi^j_m$ we shall be dealing with.

Let us consider, first of all, a highest weight field
$\Phi^j_j$. The highest weight condition implies that
the OPE's of $\Phi^j_j$ with the raising currents $j^+$
and $J^+$ must vanish. By inspecting the realization of
these currents in eq. (\ref{cincuenta}), one
immediately reaches the conclusion that in the
expression of $\Phi^j_j$ only the fields $w$ and $\phi$
can appear. We therefore shall adopt the following
ansatz for $\Phi^j_j$ :
\beq
\Phi^j_j\,=\,w^A\,e^{iB\,\alpha_0\,\phi}\,\,,
\label{citres}
\eeq
where $A$ and $B$ are constants to be determined. There
are, actually, two conditions that $A$ and $B$ must
satisfy. The first one comes from the fact that 
$\Phi^j_j$ should have a $J^0$ charge equal to $j$ and
takes the form:
\beq
A-{B\over 2}\,=\,j\,\,.
\label{cicuatro}
\eeq
Moreover, the $L_0$ eigenvalue of $\Phi^j_j$ must be
the conformal weight $h_j$ (see eq. (\ref{vsiete})).
This requirement imposes the following condition for
$A$ and $B$:
\beq
A\,+\,{B(B+1)\over 2(2k+3)}\,=\,h_j\,\,.
\label{cicinco}
\eeq
Eliminating $A$ of eqs. (\ref{cicuatro}) and 
(\ref{cicinco}), one gets a quadratic equation for $B$
which has two solutions. One of these solutions is
$A=0$, $B=-2j$, which corresponds to:
\beq
\Phi^j_j\,=\,e^{-2ij\,\alpha_0\,\phi}\,.
\label{ciseis}
\eeq
By acting on the field (\ref{ciseis}) with the lowering
operators $j^-$ and $J^-$, one can obtain the other
members $\Phi^j_m$ of the field multiplet. The result
is:
\beq
\Phi^j_m\,=\,\cases{\chi^{j-m}\,e^{-2ij\alpha_0\,\phi}
                     &if $j-m\in \ZZ$\cr\cr
                    \chi^{j-m-{1\over 2}}\,\psi\,
                    e^{-2ij\alpha_0\,\phi}
                    &if $j-m\in \ZZ\,+{1\over 2}\,$.}
\label{cisiete}
\eeq
The second solution of eqs. (\ref{cicuatro}) and 
(\ref{cicinco}) is $A=2j-k-1$ and $B=2j-2(k+1)$. This
solution corresponds to a second conjugate
representation of the highest weight field:
\beq
\tilde \Phi_j^j\,=\,w^{2j+s}\,\,
e^{2i(j+s)\alpha_0\,\phi}\,,
\label{ciocho}
\eeq
where $s\,=\,-k-1$. By successive application of the
currents $j^-$ and $J^-$, one can generate other
components of the conjugate multiplet of primary
fields. In general, the expressions of the 
$\tilde \Phi_{m}^j$ are increasingly complicated as
$m$ is decreased. To illustrate this point let us write
down the expression of the conjugate field for
$m=j-{1\over 2}$:
\beq
\tilde \Phi_{j-1/2}^j\,=\,{1\over 2j}\,[\,
(2j+s)\,\bar\psi\,w^{2j+s-1}\,-\,s\,w^{2j+s}\,\psi\,]\,
e^{2i(j+s)\,\alpha_0\phi}\,.
\label{cinueve}
\eeq
Taking $j=0$ in eq. (\ref{ciocho}), we get a conjugate
representation of the unit operator:
\beq
\tilde I\,=\,\tilde
\Phi_{0}^0\,=\,w^s\,e^{2is\alpha_0\phi}\,.
\label{sesenta}
\eeq
The expression (\ref{sesenta}) of the conjugate
identity fixes the charge asymmetry of the Fock space
metric of our free field realization. Indeed, the
condition that the expectation value of $\tilde I$ be
non-vanishing imposes a series of selection rules that
the non-zero correlators of the theory must
satisfy. Let us imagine
that we are computing the expectation value 
$<\,\prod_{i}\,O_i\,>$, where $O_i$ are general operators of
the form 
$O_i\,=\,w^{n_i}\,\chi^{m_i}\,e^{i\alpha_i\phi}\,$. Calling 
$N(w)\,=\,\sum_i\,n_i$ and $N(\chi)\,=\,\sum_i\,m_i$, one gets
the following conditions:
\bear
&&N(w)\,-\,N(\chi)\,=\,s\rc
&&\sum_i\,\alpha_i\,=\,2\alpha_0 s\,.\label{suno}\\
\nonumber
\eear

According to the standard method of the Coulomb gas
representations, the conformal blocks of
the theory can be obtained as expectation values of
products of the fields, both in the representation 
(\ref{cisiete}) and in its conjugate. The fulfillment
of the selection rules (\ref{suno}) is, in general,
achieved by the insertion of a power of the screening
charge operator $Q$ which, in our case, is given by:
\beq
Q\,=\,\oint\,dz\,
(\,\bar\psi(z)\,-\,w(z)\psi(z)\,)\,
e^{i\alpha_0\phi(z)}\,.
\label{sdos}
\eeq

Let us illustrate how our formalism works for the
two-point function. It can be easily seen that the
conditions (\ref{suno}) can be satisfied by considering
the expectation value of the product of a field
(\ref{cisiete}) and its conjugate, 
without the insertion of the screening
charge $Q$. For example, in the case of the highest
weight primary vectors, the expectation value to be
computed is:
\beq
<\,\Phi^{j}_{-j}(z_1)\,\tilde\Phi^{j}_{j}(z_2)\,>\,=\,
<\,[\chi(z_1)]^{2j}\,e^{-2ij\alpha_0\phi(z_1)}\,
[w(z_2)]^{2j+s}\,e^{2i(j+s)\alpha_0\phi(z_2)}\,>\,,
\label{stres}
\eeq
and one can prove by inspection that eq. (\ref{suno})
is satisfied. Moreover, by applying Wick's theorem, one
can write:
\beq
<\,\Phi^{j}_{-j}(z_1)\,\tilde\Phi^{j}_{j}(z_2)\,>\,=\,
{C\over (z_1-z_2)^{2h_j}}\,,
\label{scuatro}
\eeq
where $h_j$ is given in eq. (\ref{vsiete}) and $C$ is a
constant proportional to  the expectation value of
$\tilde I$. 

The four-point conformal blocks of the model can be
represented as  correlators of the form 
$<\Phi^{j_1}_{m_1}(z_1)\,\Phi^{j_2}_{m_2}(z_2)\,
\Phi^{j_3}_{m_3}(z_3)\,\tilde\Phi^{j_4}_{m_4}(z_4)\,
Q^n\,>$. The number $n$ of screening charges can be
easily determined from the second condition 
(\ref{suno}). Indeed, one can immediately demonstrate
that only when 
$n\,=\,2\,(\,j_1\,+\,j_2\,+\,j_3\,-\,j_4\,)$ this
correlator is non-vanishing. In order to study the
analytical structure of these blocks we shall
concentrate our efforts in the analysis of the quantity:
\beq
 I(z)\,\equiv\,
<\,\Phi^{j_1}_{-j_1}(0)\,\Phi^{j_2}_{j_2}(z)\,
\Phi^{j_2}_{-j_2}(1)\,\tilde\Phi^{j_1}_{j_1}(\infty)\,
Q^{4j_2}\,>\,.
\label{scinco}
\eeq
We shall assume that the four representations involved
in eq. (\ref{scinco}) are even. From the expressions of
the primary fields and the screening charge, one can
obtain the explicit form of $I(z)$:
\beq
I(z)\,=\,
\prod_{i=1}^{n}\,\,\oint_{C_i}\,\,d\tau_i\,
\lambda(z,\{\tau_i\})\,\eta(\{\tau_i\})\,,
\label{sseis}
\eeq
where $n=4j_2$, $\lambda(z,\{\tau_i\})$ is the part of
the correlator that corresponds to the field $\phi$,
namely:
\bear
\lambda(z,\{\tau_i\})\,=\,
<\,e^{-2ij_1\alpha_0\,\phi(0)}&&\,e^{-2ij_2\alpha_0\,\phi(z)}\,
e^{-2ij_2\alpha_0\,\phi(1)}\,
e^{2i(s+j_1)\alpha_0\,\phi(\infty)}\,\times\rc
&&\times e^{i\alpha_0\,\phi(\tau_1)}\cdots
 e^{i\alpha_0\,\phi(\tau_n)}\,>\,,\label{ssiete}\\
\nonumber
\eear
and the function $\eta(\{\tau_i\})$ contains the
contribution of the fields $w$, $\chi$, $\psi$ and
$\bar\psi$. It is not difficult to prove that the
non-vanishing contributions to $\eta(\{\tau_i\})$ are
of the form:
\bear
\eta(\{\tau_i\})\,=&&\,(-1)^{2j_2}\,
<\,(\chi(0))^{2j_1}\,(\chi(1))^{2j_2}\,
(w(\infty))^{2j_1+s}\,w(\tau_1)\,\cdots\,
w(\tau_{2j_2})\,\,>\times\rc\rc
&&\times\,<\,\psi(\tau_1)\cdots\psi(\tau_{2j_2})\,
\bar\psi(\tau_{2j_2+1})\cdots\bar\psi(\tau_{4j_2})\,>\,+\,
{\rm permutations.}\rc
\label{socho}
\eear
\begin{figure}
\centerline{\hskip-.4in \epsffile{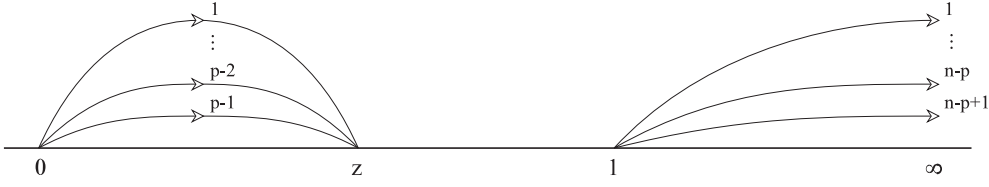}}
\caption{Contours of integration needed to represent 
$I_p(z)$.}
\label{contours}
\end{figure}

Up to now we have not specified the contours of
integration appearing in eq. (\ref{sseis}). We shall
use the canonical set of contours that give rise to the
s-channel conformal blocks (see figure \ref{contours}).
We shall take the first $n-p+1$ integrals along a path
lying on the real axis and joining the points $\tau=1$
and
$\tau=\infty$. The remaining $p-1$ integrals will be
taken along the segment $(0,z)$. Relabeling
appropriately the integration variables, the $p^{th}$
conformal block can be written as:
\bear
I_p(z)\,=\,
\int_1^{\infty}\,du_1\cdots\int_1^{u_{n-p}}\,
du_{n-p+1}\int_0^{z}\,dv_1\cdots\int_0^{v_{p-2}}\,dv_{p-1}\,
\lambda_p(z,\{u_i\},\{v_i\})\,\eta_p(\{u_i\},\{v_i\}).
\rc
\label{snueve}
\eear
In eq. (\ref{snueve}), the quantities 
$\lambda_p(z,\{u_i\},\{v_i\})$
and $\eta_p(\{u_i\},\{v_i\})$ are, respectively,  
the functions 
$\lambda(z,\{\tau_i\})$ and  
$\eta(\{\tau_i\})$ after the
relabelling of variables introduced above. 
By applying Wick's
theorem to the vacuum expectation value
(\ref{ssiete}), one can
readily  prove that  $\lambda_p(z,\{u_i\},\{v_i\})$ 
is given by:
\bear
\lambda_p(z,\{u_i\},\{v_i\})\,=&&\,z^{8j_1j_2\rho}\,
(1-z)^{8j_2^2\rho}\,
\prod_{i=1}^{n-p+1}\,u_i^a\,(u_i-z)^b\,(u_i-1)^b\,
\prod_{i<j}(u_i-u_j)^{2\rho}
\times\rc
&&\times
\prod_{i=1}^{p-1}\,v_i^a\,(z-v_i)^b\,(1-v_i)^b
\prod_{i<j}(v_i-v_j)^{2\rho}
\,\prod_{i=1}^{n-p+1}\,\prod_{j=1}^{p-1}
(u_i-v_j)^{2\rho},\rc
\label{setenta}
\eear
where $\rho\,=\,\alpha_0^2/2$, $a\,=\,-2j_1\alpha_0^2$
and $b\,=\,-2j_2\alpha_0^2$. It is not difficult to
obtain the non-analytical behaviour of the blocks
around the point $z=0$. This behaviour is of the form:
\beq
I_p(z)\,\sim\,N_p\,z^{\gamma_p}\,,
\label{stuno}
\eeq
where $N_p$ and $\gamma_p$ are constants. The latter
can be written as a difference of conformal weights of
the form:
\beq
\gamma_p\,=\,h_{j_3}\,-\,
h_{j_1}\,-\,h_{j_2}\,.
\label{stdos}
\eeq
The isospin $j_3$ has the interpretation of the isospin
of the s-channel intermediate state. Its expression as
a function of $p$ is:
\beq
j_3\,=\,j_1\,+\,j_2\,+{1\,-\,p\over 2}\,.
\label{sttres}
\eeq
Notice that as $p=1,\cdots, 4j_2+1$ the values taken by
$j_3$ are $j_1-j_2, j_1-j_2+{1\over 2},\cdots,
j_1+j_2$, in agreement with the Clebsch-Gordan
decomposition (\ref{trece}). 

The physical correlation functions, which we shall
denote by $G(z\,,\,\bar z)$, can be obtained by
combining holomorphic and antiholomorphic blocks in a
monodromy invariant way:
\beq
G(z\,,\,\bar z)\,=\,\sum_p\,X_p\,|\,I_p(z)\,|^2\,.
\label{stcuatro}
\eeq
The coefficients $X_p$ have been computed in ref.
\cite{DF}. The leading $z\rightarrow 0$ behaviour of 
$G(z\,,\,\bar z)$ can be obtained by combining eqs. 
(\ref{stuno}) and (\ref{stcuatro}):
\beq
G(\,z\,,\,\bar z\,)\,\sim\,\sum_p\,\,\Bigl[\,
{S_p\over |z|^{2\,(h_{j_1}+h_{j_2}
-h_{j_3})}}\,+\,O(z)\,\Bigr]\,,
\label{stcinco}
\eeq
where the constants $S_p$ are given by:
\beq
S_p\,=\,X_p\,(N_p)^2\,.
\label{stseis}
\eeq
The quantities $S_p$ are related to the structure
constants of the operator product algebra of the model.
These constants, which we shall denote by 
$D_{j_1,m_1;j_2,m_2}^{j_3,m_3}$, appear in the leading
terms of the OPE's of the primary fields, namely:
\beq
\Phi_{m_1}^{j_1}(z_1,\bar z_1)\,\Phi_{m_2}^{j_2}(z_2,\bar z_2)\,=\,
\sum_{j_3,m_3}\,D_{j_1,m_1;j_2,m_2}^{j_3,m_3}\,\,\Bigl[
\,{\Phi_{m_3}^{j_3}(z_2,\bar z_2)\over 
|z_1\,-z_2|^{2(h_{j_1}\,+\,h_{j_2}\,-\,h_{j_3})}}\,+\,
O(z_1\,-z_2)\,\Bigr].
\label{stsiete}
\eeq
The two-point functions of the theory are normalized
as:
\beq
<\,\Phi_{m_1}^{j_1}(z_1,\bar z_1)\,
\Phi_{m_2}^{j_2}(z_2,\bar z_2)\,>
\,=\,(-1)^{\sigma(j_1,m_1)}\,\,\,
{\delta_{j_1,j_2}\,\delta_{m_1,-m_2}\over
|z_1\,-z_2|^{4h_{j_1}}}\,,
\label{stocho}
\eeq
where $\sigma(j,m)$ is $0$($1$) if the state 
$|j, m>$ has positive(negative) norm. Therefore, the
structure constants must satisfy the constraint:
\beq
D_{j_1,m_1;j_1,-m_1}^{0,0}\,=\,
(-1)^{\sigma(j_1,m_1)}\,.
\label{stnueve}
\eeq
In order to relate the quantities $S_p$ of eq. 
(\ref{stseis}) to the structure constants
(\ref{stnueve}), let us use the OPE's (\ref{stsiete}) 
in the correlator $G(\,z\,,\,\bar z\,)$. The result one
gets is:
\beq
G(\,z\,,\,\bar z\,)\,\sim\,\sum_{j_3,m_3}\,\,
(-1)^{\sigma(j_3,m_3)}\,\,\Bigl[\,
{[\,D_{j_1,j_1;j_2,-j_2}^{j_3,m_3}\,]^2\over
|z|^{2\,(h_{j_1}+h_{j_2}
-h_{j_3})}}+\,O(z)\,\Bigr]\,,
\label{ochenta}
\eeq
from which one we have the identification:
\beq
(-1)^{\sigma(j_3,m_3)}\,\,
[\,D_{j_1,j_1;j_2,-j_2}^{j_3,m_3}\,]^2\,\,\sim
\,\,S_p\,.
\label{ouno}
\eeq
Using (\ref{ouno}) it is possible to obtain the
structure constants from our free field formalism
\cite{osp}. Let us introduce the functions $\lambda(j)$
and 
${\cal P}(j)$. The former is defined as:
\beq
\lambda(j)\,\equiv\,
{\Gamma({j\over 2}\,+\,j\rho\,-\,[{j\over 2}])\over
\Gamma({j\over 2}\,-\,j\rho\,-\,[{j\over 2}])}\,\,,
\label{odos}
\eeq
while ${\cal P}(j)$ is given by:
\beq
{\cal P}(j)\,\equiv\,\prod_{i=1}^{j}\,\lambda(i)\,=\,
\prod_{i=1}^{j}\,
{\Gamma({i\over 2}\,+\,i\rho\,-\,[{i\over 2}])\over
\Gamma({i\over 2}\,-\,i\rho\,-\,[{i\over 2}])}\,\,.
\label{otres}
\eeq
Let us also introduce the Clebsch-Gordan coefficients
corresponding to the tensor product decomposition 
(\ref{trece}):
\beq
|\,j_3,m_3\,>\,=\,\sum_{m_1,m_2}\,
C_{j_1,m_1;j_2,m_2}^{j_3,m_3}\,\,\,\,
|j_1,m_1>\otimes\,|j_2,m_2>\,\,.
\label{ocuatro}
\eeq
In terms of the quantities defined above, the structure
constants can be written as \cite{osp}:
\bear
\Bigl[\,D_{j_1,m_1;j_2,m_2}^{j_3,m_3}\,\Bigr]^2\,=&&\,
\Bigl[\,C_{j_1,m_1;j_2,m_2}^{j_3,m_3}\,\Bigr]^4\,
\lambda(1)\,\,
{\cal P}^2(2j_1+2j_2+2j_3+1)\,
\times\rc\rc
&&\times
\,\,\prod_{i=1}^{3}\,\,
{\lambda(4j_i+1)\,
{\cal P}^{2}(2j_1+2j_2+2j_3-4j_i)\over 
{\cal P}^2(4j_i+1)}\,\,. \label{ocinco}\\
\nonumber
\eear
By studying the conditions under which the right-hand
side of eq. (\ref{ocinco}) is non-vanishing we can
obtain the fusion rules of the model. First of all, it
is easy to verify that those fields with isospin 
$j\,\le\,k/2$ close under multiplication. Actually, a 
detailed study of eq. (\ref{ocinco}) (see ref.
\cite{osp}) leads to the fusion rule:
\beq
[j_1]\,\times\,[j_2]\,=\,
\sum_{{j_3=|j_1-j_2|\atop}\atop 2(j_3-j_1-j_2)\,\in\,\ZZ}
^{{\rm min}\,(\,j_1+j_2\,,\,k+{1\over 2}-j_1-j_2\,)}
\,\,\,[j_3]\,\,,
\label{oseis}
\eeq
which can be compared with the composition law of the
finite algebra (eq. (\ref{trece})).

\setcounter{equation}{0}
\section{ Conclusions and final remarks}
\label{ospcon}

In previous sections we have reviewed a series of
results which have been recently obtained for the CFT
based on the osp${(1\vert 2)}$ affine Lie
superalgebra. The global picture emerging from these
results is that the osp${(1\vert 2)}$ current
algebra is a perfectly solvable rational CFT. In order
to complete this picture it would be desirable to study
some other aspects of the theory. Let us mention some
of them. First of all, one should explore the
possibility of building a CFT for the admissible
representations, with fractional levels and isospins
given by eq. (\ref{tocho}). The fusion rules for these
representations have been determined in ref. \cite{null}
from the null vector decoupling conditions. 

Coming back to
the case in which the isospin is integer or half-integer
and the level $k$ is a non-negative integer, it is
interesting to study the crossing symmetry of the
conformal blocks of the theory. One can employ
\cite{dual} with this purpose the free field
representation of section \ref{ospfree}. The behaviour of
the correlator of the theory under exchange symmetry,
\ie\ under the braiding and fusion operations, should be
determined by a quantum deformation of the universal
enveloping algebra of  osp${(1\vert 2)}$. Moreover, this
behaviour could be used to define invariants for
three-manifolds. The corresponding Chern-Simons theory,
whose states are in one-to-one correspondence with the
conformal blocks of the two-dimensional model,  allows
to define knot invariants. We have recently found
\cite{dual} the relation of these invariants with the
su(2) knot polynomials. Let us finally mention that,
with these results at hand, one could also study the
integrable deformation of the  osp${(1\vert 2)}$ CFT
with the hope of finding new solvable massive field
theories in two dimensions.

\setcounter{equation}{0}
\section{ Acknowledgements}
\label{ospack}

Two of us (JMSS and AVR) would like to thank the
organizers of the workshop ``Trends in Theoretical
Physics" for their warm hospitality at La Plata.  This
work was supported in part by DGICYT under grant
PB93-0344,  by CICYT under grant  AEN96-1673 and by the
European Union TMR grant ERBFMRXCT960012.


\begin{thebibliography}{99}

\bibitem{Review} For a review see J. Fuchs, {\sl
``Affine Lie algebras and quantum groups"}, Cambridge
University Press, 1992 and  S. Ketov, {\it ``Conformal
Field Theory"}, World Scientific, Singapore(1995).




\bibitem{bershadsky}M. Bershadsky and H.
Ooguri, {\sl \pl} {\bf B229} (1989) 374.


\bibitem{poly}A. M. Polyakov and A. B.
Zamolodchikov, {\sl \mpl} {\bf A3} (1988) 1213.


\bibitem{yu} J. B. Fan and M. Yu, {\sl ``G/G Gauged
Supergroup Valued WZNW Field Theory"}, Academia Sinica
preprint AS-ITP-93-22, hep-th/9304123.


\bibitem{Ennes}I. P. Ennes, J. M. Isidro and A. V.
Ramallo, {\sl \ijmp} {\bf A11} (1996)2379.


\bibitem{Pais}A. Pais and V.
Rittenberg, {\sl \jmp} {\bf 16}(1975) 2063;  M.
Scheunert, W. Nahn and  V.
Rittenberg, {\sl \jmp} {\bf 18}(1977)155.



\bibitem{KW}V. Kac and M.
Wakimoto, {\sl \pnas} {\bf 85} (1988)4956.




\bibitem{yudos}J. B. Fan and M. Yu, {\sl ``Modules over
affine Lie superalgebras"} , Academia Sinica preprint
AS-ITP-93-14, hep-th/9304122.



\bibitem{null} I. P. Ennes and A. V. Ramallo, 
{\sl Fusion rules and singular vectors of the 
osp${(1\vert 2)}$ current algebra}, Santiago preprint 
US-FT-12/97, hep-th/9704065, to appear in {\sl Nuclear
Physics B}. 




\bibitem{Gordon} B. Gordon, {\sl \qjm} {\bf
12}(1961)285.


\bibitem{osp}I. P. Ennes, A. V. Ramallo and J. M. Sanchez
de Santos, {\sl \pl} {\bf B389}(1996)485;  
{\sl \np} {\bf B491 [PM]} (1997) 574.




\bibitem{DF}Vl.S.Dotsenko and V. A. Fateev
{\sl \np} {\bf B240}(1984)312;
{\sl\np} {\bf B251}(1985)691; {\sl \pl} {\bf B154}
(1985)291.


\bibitem{dual}I. P. Ennes, P. Ramadevi, A. V. Ramallo
and J. M. Sanchez de Santos, ``Duality in 
osp${(1\vert 2)}$ Conformal Field Theory and link
invariants", preprint in preparation.



 









\end{thebibliography}
\end{document}